# A simple superluminal but not physical motion

**Nilton Penha**
Departamento de Física, Universidade Federal de Minas Gerais, Brasil.
nilton.penha@gmail.com

**Bernhard Rothenstein**
Politehnica University of Timisoara, Physics Department, Timisoara, Romania.
brothenstein@gmail.com

**Abstract.** We discuss a superluminal unphysical motion which, we believe, has a high pedagogical potential. It highlights the physics behind the concept of simultaneity in special relativity and illustrates the non-physical character of superluminal speeds offering a rewarding exercise in handling the Minkowski space-time diagram.

Consider a three-dimensional spacetime (two spatial dimensions + one time dimension) and two inertial frames $K(x,y,ct)$ and $K'(x',y',ct')$ according to which a given event $E$ has respectively the following formal representations: $E(x,y,ct)$ and $E'(x',y',ct')$. The coordinates $x,y,z$ ($x',y',z'$) are Cartesian coordinates which account for the event spatial position and coordinate $ct(ct')$ accounts for its time position; $c$ is the absolute speed of light in vacuum, taken as constant according to the second postulate of Einstein's special theory of relativity, the first postulate being that "all physics laws are the same for all inertial reference frames".

Suppose that $K$ and $K'$ are moving away from each other with a constant speed $V$. Because of the isotropy of space, one can orient both *x-axis* and *x'-axis* along the direction of speed $V$; *y-axis* and *y'-axis* are taken parallel to each other. For convenience consider also that both frame origins coincide, ie, $(x,y,ct) = (0,0,0) = (x',y',ct')$. This will be the spacetime origin point.

In each of the inertial frames it is convenient to imagine a standard lattice of stationary observers as small as can be, each one with a standard clock, being all the clocks alike and stationary. It is common among physicists to refer figuratively to such clocks as wristwatches. It is also convenient to have all the clocks synchronized. To achieve synchronization, one can use Einstein's prescription.

Take the inertial frame $K$. Initially all the clocks are stopped. And then one should choose a master clock, usually the one at the spatial origin of inertial frame in question. The master clock should be set to start running at time $ct_o$ *(usually zero)* and all the others set to start running at time $ct = ct_0 + r$ where $r$ is the spatial distance from the master clock. Then a light source previously placed at the master clock lattice site emits a pulse at time $ct_0$. The pulse propagates through the lattice and triggers off each one of the clocks which start reading $ct = ct_o + r$ as previously settled. Once all the clocks are running they are all synchronized. But, while this process is not finished, all the lattice clocks inside a circle of radius

$$r = \sqrt{x^2 + y^2} = ct \qquad (1)$$

are synchronized. Those outside the circle are not synchronized yet; actually they are all stopped, by construction, waiting for the light pulse. So one has a circular frontier between the synchronized and unsynchronized clocks regions; such frontier propagates outwards, with constant speed $c$, from the master clock, which is at the spatial origin. All events on such outgoing circle, which coincides with the light wavefront, are considered simultaneous to observers at rest in $K$. See Figure 1 for the scenario of our discussion here.

A similar synchronization procedure should be done in the inertial frame $K'$ which is moving away from $K$ with speed $V$. A light pulse emitted by a source in $K'$ master clock propagates through a standard lattice of observers and their wristwatches. In the same manner as in $K$, the pulse which propagates through the referred lattice should trigger off each one of the existing clocks (wristwatches) which start reading $ct' = ct_0' + r'$ as previously settled. Once all the clocks are running they are all synchronized. But, in the same way as in *K*, while this process is not finished, all the lattice clocks inside a circle of radius

$$r' = \sqrt{x'^2 + y'^2} = ct' \qquad (2)$$

are synchronized. Those outside the circle are not synchronized; actually they are all stopped, by construction, waiting for the light pulse. So one has a circular frontier between the synchronized and unsynchronized clocks regions; such frontier propagates outwards, with constant speed *c*, from the master clock, which is at the spatial origin. All events on such outgoing circle, which coincides with the light wavefront, are considered simultaneous to observers at rest in *K*. See Figure 1 for the scenario of our discussion here where one should, in this case, replace unprimed for primed parameters. The first postulate of relativity ensures that physics is the same in different inertial frames.

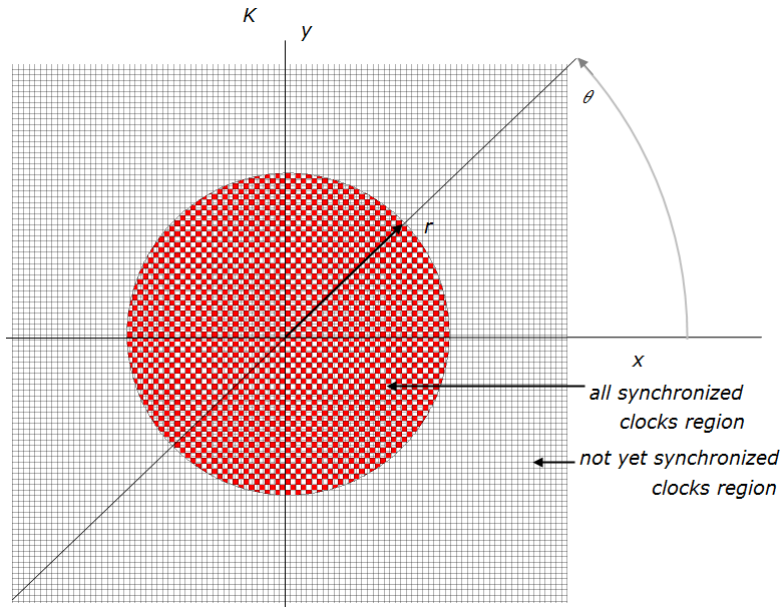

**Figure 1** –In Einstein synchronization procedure one has a circular frontier between the synchronized clocks region (inside the circle) and the not yet synchronized clocks region (outside the circle) according to observers at rest in frame *K*.

Now let these synchronization processes start at the exact moment when the origin of both inertial reference frames is the same: $(x_0, y_0, ct_0) = (0,0,0) = (x_0', y_0', ct_0')$. The two master clocks are face to face, at this same origin, reading $ct_0 = 0 = ct_0'$, by construction, when a light source emits a pulse at the common spatial origin. The stationary observers in *K* and *K'* "see" circular wavefronts centered at their respective master clock, in terms of their ***own spacetime coordinates***. All the clocks at rest in *K*, which are instantly on a circular wavefront and inside the circle, read the same time $ct = r$. All the clocks at rest in *K'*, which are on a circular wavefront and inside the circle, read the same time $ct' = r'$. The relation between spacetime coordinates in *K* and *K'* is given by Lorentz Transformations.

One can investigate now how the simultaneity in *K'* is related to the simultaneity in *K*.

Let *E'*(*x'*, *y'*, *ct'*) be a set of simultaneous (*fixed ct'*) events according to observers at rest in *K'*, in Cartesian coordinates, and *E'*( *r'cos* ', *r'sin* ', *ct'*) in polar coordinates as well. By applying appropriately Lorentz transformations to the simultaneous events spacetime coordinates in *K'*, one gets how these same events are "seen" by stationary observers in *K*: *E*(*x, y, ct*) where

$$x = \gamma (x' + \beta \, ct') \qquad (3)$$

$$y = y' \tag{4}$$
$$ct = \gamma (ct' + \beta x') \tag{5}$$

or $E(r\cos\theta, r\sin\theta, ct)$, where

$$r\cos\theta = \gamma (r'\cos\theta' + \beta ct'), \tag{6}$$
$$r\sin\theta = r'\sin\theta', \tag{7}$$
$$ct = \gamma (ct' + \beta r'\cos\theta'), \tag{8}$$

using standard relativistic notations $\beta = V/c$ and $\gamma = 1/\sqrt{1-\beta^2}$. The polar coordinates $r$, in $K$ are explicitly expressed as

$$r = \gamma\, r' \sqrt{(\cos\theta' + \beta \frac{ct'}{r'})^2 + \frac{(\sin\theta')^2}{\gamma^2}}, \tag{9}$$

$$\theta = \arctan\left(\frac{\sin\theta'}{\gamma\left(\cos\theta' + \beta \frac{ct'}{r'}\right)}\right). \tag{10}$$

Recall that a pulse is emitted at $ct_0 = 0 = ct_0'$ by the light source at the common spatial origins. A circular wavefront propagates radially outward from the origin, at speed $c$, which is constant, no matter the direction it goes, according to second postulate of special relativity. According to observers at rest in $K$ the circular wavefront has a radius $r = ct > 0$ and according to observers at rest in $K'$ the circular wavefront has radius $r' = ct' > 0$.

To stationary observers in $K$, all the points on a given circular wavefront ($r = ct = constant$) constitute the geometrical locus of simultaneous events $E(x, y, r)$ or $E(r\cos\theta, r\sin\theta, r)$. Obviously, $x$ and $y$, on the wavefront satisfy the equation (1).

To stationary observers in $K'$, all the points on a given circular wavefront ($r' = ct' = constant$) constitute the geometrical locus of simultaneous events $E'(x', y', r')$ or $E'(r'\cos\theta', r'\sin\theta', r')$. Obviously, $x'$ and $y'$, on the wavefront satisfy the equation (2).

Performing Lorentz Transformations on $K'$ spacetime coordinates reveals that the shape of frontier between the synchronized and the unsynchronized clocks region (such frontier is circular to $K'$) appears as an ellipse to the "stationary" observer in $K$, as shown in Figure 2, for some values of $\beta$ and a fixed $r' = ct' = 1$ (arbitrary unit). The corresponding stationary $K$ spacetime coordinates are given by

$$x = \gamma\, r'(\cos\theta' + \beta) = r\cos\theta, \tag{11}$$
$$y = r'\sin\theta' = r\sin\theta, \tag{12}$$
$$ct = \gamma\, r'(1 + \beta \cos\theta'). \tag{13}$$

or

$$r = \gamma\, r'(1 + \beta \cos\theta'), \tag{14}$$

$$\theta = \arctan\left(\frac{\sin\theta'}{\gamma(\cos\theta' + \beta)}\right), \quad \cos\theta = \frac{\beta + \cos\theta'}{1 + \beta \cos\theta'}, \quad \sin\theta = \frac{\sin\theta'}{\gamma(1 + \beta \cos\theta')}, \tag{15}$$

$$ct = r \tag{16}$$

One easily verifies that

$$(ct')^2 - (x'^2 + y'^2) = (ct)^2 - (x^2 + y^2) = 0, \tag{17}$$

as it should be, since light travels on the lightcone surface where invariant spacetime interval is zero.

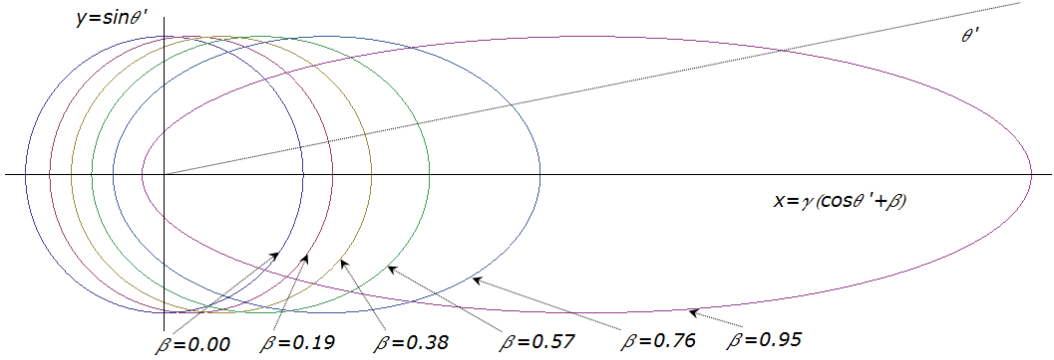

**Figure 2** – Circular geometrical locus in the inertial moving frame K' are "seen" in the inertial stationary frame K as ellipses shown for different values of β and fixed $r' = ct' = 1$ (arbitrary unit).

Consider now two of the simultaneous point events, $E_1'(x_1', y_1', ct_1') = E_1'(r_1'\cos\theta_1', r_1'\sin\theta_1', ct_1')$ and $E_2'(x_2', y_2', ct_2') = E_2'(r_2'\cos\theta_2', r_2'\sin\theta_2', ct_2')$, on a circle of radius $r' = r_1' = ct_1' = r_2' = ct_2' = ct' = 1$ (arbitrary unit) with polar angles $\theta_1' = 3\pi/4$ and $\theta_2' = \pi/4$:

$$E_1'(-\frac{1}{\sqrt{2}}, +\frac{1}{\sqrt{2}}, 1) \tag{18}$$

and

$$E_2'(+\frac{1}{\sqrt{2}}, +\frac{1}{\sqrt{2}}, 1). \tag{19}$$

These same two events are labeled in K as $E_1(x_1, y_1, ct_1) = E_1(r_1\cos\theta_1, r_1\sin\theta_1, ct_1)$ and $E_2(x_2, y_2, ct_2) = E_2(r_2\cos\theta_2, r_2\sin\theta_2, ct_2)$,

$$E_1(\gamma(\beta - \frac{1}{\sqrt{2}}), \frac{1}{\sqrt{2}}, \gamma(1 - \beta\frac{1}{\sqrt{2}})) \tag{20}$$

and

$$E_2(\gamma(\beta + \frac{1}{\sqrt{2}}), \frac{1}{\sqrt{2}}, \gamma(1 + \beta\frac{1}{\sqrt{2}})). \tag{21}$$

where the corresponding polar angles are

$$\theta_1 = \pi + \arctan(\frac{1}{\gamma(-1+\sqrt{2}\beta)}) \tag{22}$$

$$\theta_2 = \arctan(\frac{1}{\gamma(1+\sqrt{2}\beta)}) \tag{23}$$

We immediately see that $ct_1$ is not equal to $ct_2$, leading to the conclusion that $E_1$ and $E_2$ are not simultaneous [1] under the viewpoint of a stationary observer in K. In the inertial frame K $E_1$ belongs to a circle of radius

$$r_1 = \gamma\,(1-\beta\,\frac{1}{\sqrt{2}}) = ct_1 \tag{21}$$

and $E_2$ belongs to a different circle of radius

$$r_2 = \gamma\,(1+\beta\,\frac{1}{\sqrt{2}}) = ct_2 > r_1. \tag{22}$$

A stationary observer in $K$ "sees" event $E_1$ first and then $E_2$. See Figures 3 and 4.

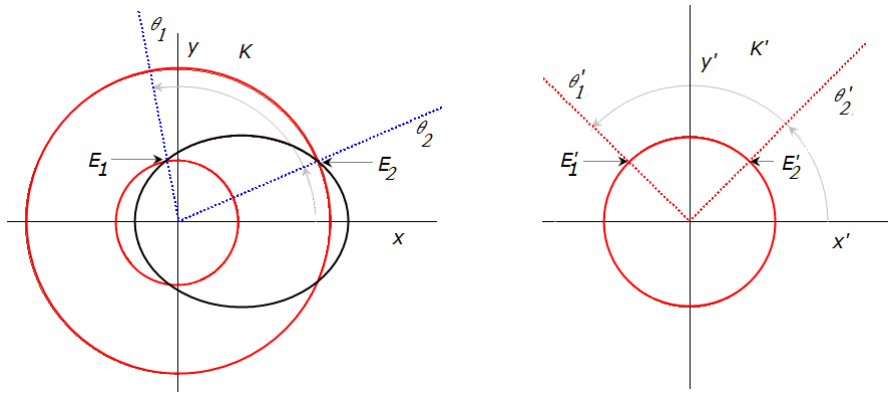

**Figure 3** – At right, a circular wavefront, in $K'$, emitted by a light source at the common origin is shown. Two simultaneous events $E_1'$ and $E_2'$ on it are chosen. At left, the circle in $K'$ is "seen" in $K$ as an ellipse containing the same two events which are now called $E_1$ and $E_2$; they are not simultaneous to stationary observers in $K$. ($\beta = 3/5$). $E_1$ and $E_2$ belong to different circular wave fronts.

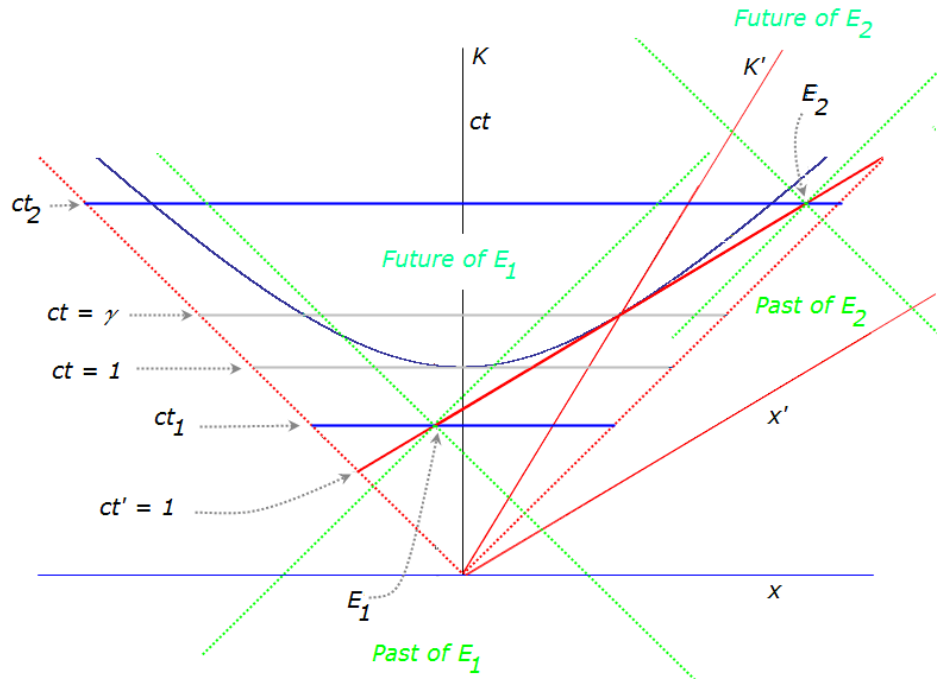

**Figure 4** – The $y = y' = 0$ Minkowski spacetime diagram. The two $E_1'$ and $E_2'$ simultaneous events for $K'$ observers are clearly not simultaneous to observers in $K$ which see first event $E_1$ and next event $E_2$. $E_1'$ and $E_1$ are one same event; $E_2'$ and $E_2$ are also another same event. Simultaneity is relative. Neither of the two events belongs to future of the other, and then they have no causal dependence.

Let us consider for a moment the quantity $(x_2-x_1)/(ct_2-ct_1)$ which has dimension of speed:

$$\frac{x_2 - x_1}{ct_2 - ct_1} = \frac{(\cos\theta_2' + \beta) - (\cos\theta_1' + \beta)}{(1 + \beta\cos\theta_2') - (1 + \beta\cos\theta_1')} = \frac{1}{\beta}. \quad (24)$$

This result does not depend on particular values of $\theta_1'$ and $\theta_2'$ and is valid for any two of the simultaneous events on any circular wavefront in $K'$. In fact if we pick expressions (11) and (13) for $x$ and $ct$, assuming $r' = ct'$ fixed, we get

$$x = u_{ap} t - \frac{ct'}{\beta\gamma}, \quad (25)$$

where the quantity $u_{ap}$ is a superluminal speed

$$u_{ap} = \frac{c^2}{V}. \quad (26)$$

See Figure 5 for a plot of $x$ versus $ct$ for $ct' = 1$ and different values of $\beta$.

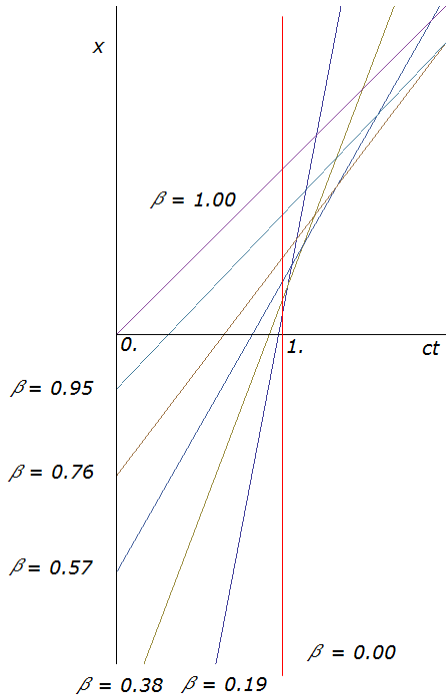

**Figure 5** – The spacetime coordinate $ct$, in $K$, is plotted against $x$, also in $K$, for different values of $\beta$ and $ct' = 1$ *(arbitrary unit)*. Notice that the slopes $(\Delta x/c\Delta t)$ are bigger than *1*, which means speeds faster than that of light.

One, usually, considers that the $u_{ap}$ is not a physical speed because the events to which it refers are not in a causal relationship to each other. Remember that $E_1'$ and $E_2'$, in the discussion above, are on the light wavefront in $K'$ and are simultaneous, with no causal dependence between them. Two events which are not in causal dependence do not belong to the future (past) of the other. In other words, they are not inside the light cone of the other. See figure 4 for an illustration of this situation in Minkowski diagram.

Now one considers a standard lattice of light sources in *K'*, that simultaneously emit each one a pulse. In Figure 6 some of the lattice elements are represented, for $y = 0 = y'$, in Minkowski spacetime diagram. The spatial sequence of eight simultaneous events appear to stationary observers in *K* in a time sequence $ct_1 < ct_2 < ct_3$ ... leading to an "effect" which propagates along *x-axis* with an apparent speed $u_{ap} = c^2/V$ which is faster than that of light. In one of his books, Rindler [2] proposes a problem with a similar scenario.

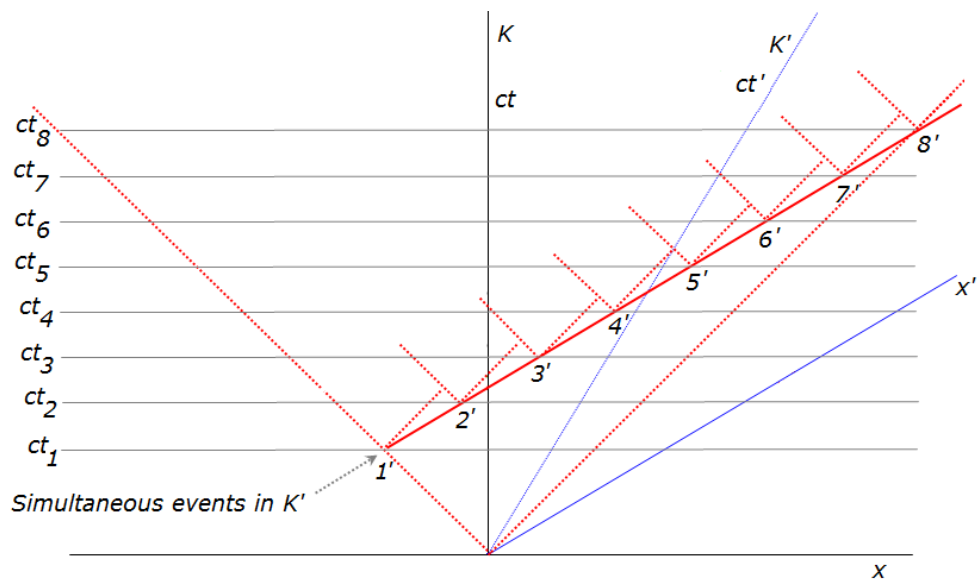

**Figure 6** – The spatial sequence of flashes emitted simultaneously in *K'* appears to stationary observers in *K* in a time sequence ($ct_1 < ct_2 < ct_3 < ct_4$ ...). The consequence of this is an "effect" which propagates along *x-axis* with an apparent speed $u_{ap} = c^2/V$ which is faster than that of light.

The relativistic notion of simultaneity makes it clear that no information can travel faster than light without throwing all our concepts of cause and effect into trouble. Since most physicists [3] still believe that cause needs to precede effect, the conclusion is that no information can be transmitted faster than the speed of light. Nevertheless, velocities greater than *c* can be observed. Suppose a lighthouse [4] that illuminates a distant building. The rotating lamp moves quite slowly, but the spot on the distant building travels at a far greater velocity. If the building were far enough away, the spot could even move faster than light. Each point along the building wall receives its own spot of light from the lighthouse, and any information travels from the lighthouse at *c*, rather than along the path of the moving spot. Such phenomena are described as the "motion of effects", and are not forbidden by relativity.

Things that supposedly travel with speed faster than light, if they exist, are called tachyons. No tachyons [5] have been definitely found and most physicists doubt their physical existence. One interesting thing about tachyons is that since they move faster than light, we would never see them approaching; they would be observed only departing.

**Conclusion**

Starting with Einstein's synchronization procedure we illustrate the physics behind the simultaneity of events and the non-physical character of superluminal speeds using the Minkowski space-time diagram.